\documentclass[twocolumn,showpacs,preprintnumbers,amsmath,amssymb,aps,prl,superscriptaddress]{revtex4-1}
\usepackage[T1]{fontenc}
\usepackage[latin1]{inputenc}
\usepackage{graphicx,xcolor}

\setcitestyle{super}

\def\change#1#2{\emph{#1} [\bgroup\color{red}#2\egroup]}

\newcommand{\beq}{\begin{equation}}
\newcommand{\eeq}{\end{equation}}

\newcommand{\beqa}{\begin{eqnarray}}
\newcommand{\eeqa}{\end{eqnarray}}

\newcommand{\Tr}[1]{\operatorname{Tr}\left[#1\right]}

\newcommand{\Eqref}[1]{Eq.~\eqref{#1}}
\newcommand{\Figref}[1]{Fig.~\ref{#1}}

\makeatother

\begin{document}

\title{Unravelling the role of inelastic tunneling into pristine and defected graphene}

\author{Mattias L. N. Palsgaard}
\affiliation{Center for Nanostructured Graphene, Dept. of Micro- and Nanotechnology, Technical University of Denmark, {\O}rsteds Plads, Bldg.~345E, DK-2800 Kongens
Lyngby, Denmark}
\author{Nick P. Andersen}
\affiliation{Center for Nanostructured Graphene, Dept. of Micro- and Nanotechnology, Technical
University of Denmark, {\O}rsteds Plads, Bldg.~345E, DK-2800 Kongens
Lyngby, Denmark}
\author{Mads~Brandbyge}
\affiliation{Center for Nanostructured Graphene, Dept. of Micro- and Nanotechnology, Technical
University of Denmark, {\O}rsteds Plads, Bldg.~345E, DK-2800 Kongens
Lyngby, Denmark}
\email{mads.brandbyge@nanotech.dtu.dk}
\date{\today}


\begin{abstract}
  We present a first principles method for calculating the inelastic electron tunneling spectroscopy (IETS) on gated
  graphene. We reproduce experiments on pristine graphene and point out the importance of including several
  phonon modes to correctly estimate the local doping from IETS. We demonstrate
  how the IETS of typical imperfections in graphene can yield characteristic
  fingerprints revealing e.g.\ adsorbate species or local buckling. Our results show
  how care is needed when interpreting STM images of defects due to suppression of the
  elastic tunneling on graphene.
\end{abstract}


\maketitle

Imperfections such as lattice defects, edges, and impurity/dopant atoms can degrade the
superb transport properties of graphene,\cite{Geim2007,Dirac,K.S.Novoselov2004,A.C.Ferrari2006} or may, if controlled, lead to new
functionality.\cite{Waveguide}
Scanning Tunneling Microscopy/Spectroscopy (STM/STS) have been used extensively to obtain
insights into the local electronic structure of graphene with atomic
resolution.\cite{Crommie,BN,PhysRevLett.104.036805,nl2041673,nl3039508} However, contrary to most STM/STS experiments where elastic tunneling plays
the dominant role, for graphene the inelastic tunneling prevails. This was clearly
demonstrated experimentally as a ``giant'' signal in the second derivative of the current
w.r.t.\ voltage obtained in Inelastic Electron Tunneling Spectroscopy (IETS) performed on
gated, pristine graphene with STM.\cite{Crommie,BN,PhysRevLett.104.036805} The pronounced inelastic features
are rooted in the electronic structure of graphene. The electrons have to enter the
Dirac-points corresponding to a finite in-plane momentum leading to weak elastic tunneling.
The IETS signal of pristine graphene has been reproduced qualitatively by \citeauthor{Wehling} considering
the change in the wavefunction decay when displacing the carbon atoms along a
selected frozen zone-boundary out-of-plane phonon.\cite{Wehling} In general, the important
role of the inelastic process complicates the interpretation of STM results on
graphene. Ideally STM images on graphene structures should be accompanied by local
STS/IETS measurements, in order to distinguish between contributions from the inelastic and elastic channel.
On the other hand, first principles calculations based on Density Functional Theory (DFT) often
provide essential unbiased insights into STM/STS/IETS experiments to help the interpretation.

In this work we present a method for DFT calculations of the STS/IETS on gated
graphene. We demonstrate its predictive power by reproducing from first principles the
features of the experimental results for the giant inelastic conductance of gated
pristine graphene.\cite{Crommie,BN,PhysRevLett.104.036805} We then provide results for IETS signals of defected graphene systems
determining the relative impact on the current of the various phonon modes. In particular
we identify inelastic fingerprints of selected defects, suggesting that IETS measurements
can be a powerful tool in the characterization of imperfect graphene. Our
analysis also illustrates how one should keep in mind the in-plane momentum conservation when
performing STM on graphene. In particular we demonstrate how defects can locally lift the suppression of elastic tunneling. The resulting increased local conductance may be misinterpreted as a high local density of states (LDOS).\\

\begin{figure}[!h]
  \centering
  \includegraphics[width=0.35\textwidth]{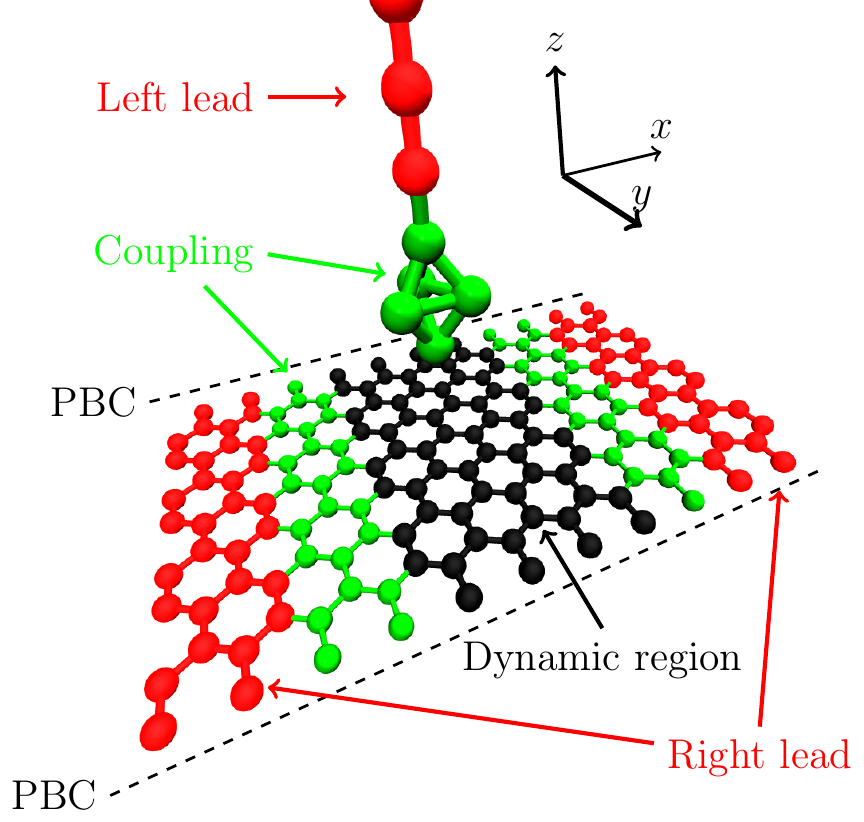} 
  \caption{(Color online) left: The system setup with semi-infinite leads (red), device region
      (green), dynamic region (black) and periodic boundary conditions along the dashed
      lines.}
  \label{Dynamic}
\end{figure}
\begin{figure}[!h]
  \centering
  \includegraphics[width=0.5\textwidth]{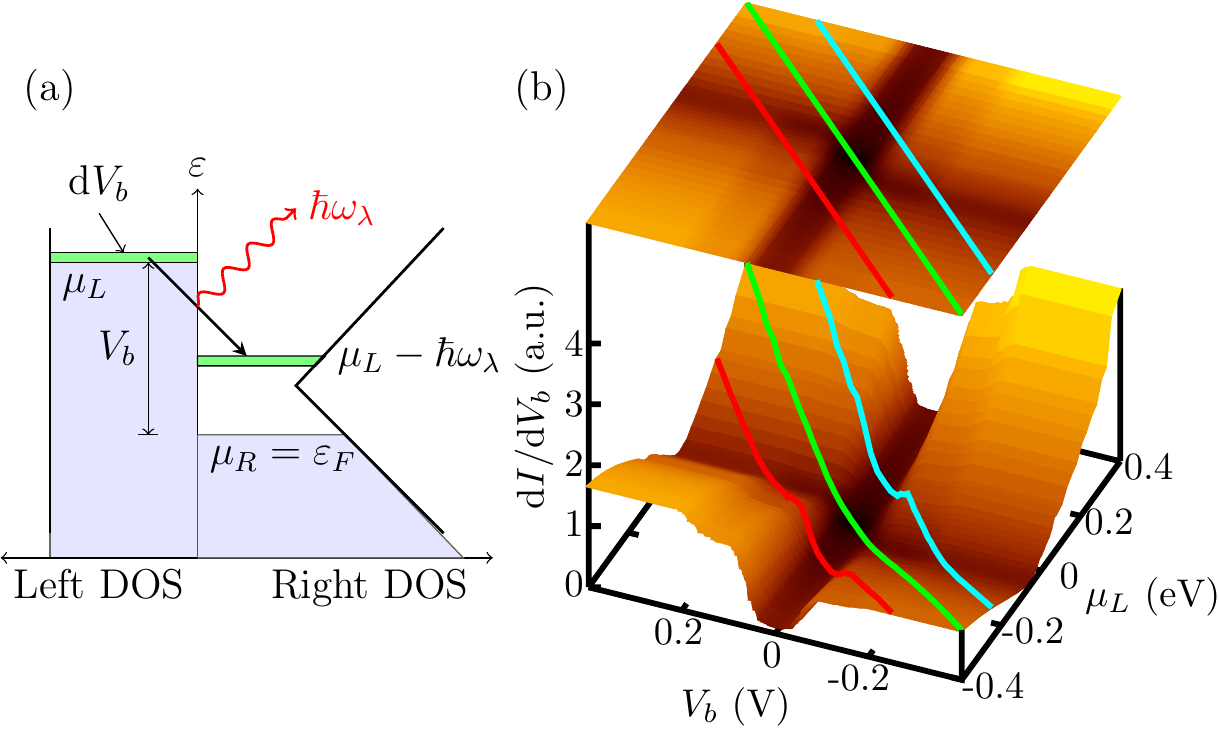} 
  \caption{(Color online) (a) Energy diagram showing the important (green) contributions of the
      left/right DOS when calculating the differential conductance for $V_b>\hbar\omega_\lambda$.
      (b) LOE differential conductance as a function of bias and tip(``left'') chemical
      potential. The lines ($\mu _L =eV_b + \varepsilon _F $) indicate the contours taken
      to include above threshold effects at a gate voltage ($\varepsilon_F$) of $0$ eV
      (green) and $\pm 0.15$ eV (blue/red).}
  \label{phase}
\end{figure}
\noindent
\textit{Method.}\quad The calculations are performed with DFT using the SIESTA/TranSIESTA
\cite{SIESTA,TranSIESTA} code and the Inelastica package for inelastic
transport.\cite{Inelastica} Our system shown in \Figref{Dynamic} is divided into device
and ``left''/``right'' leads following the standard\cite{TranSIESTA,Inelastica} transport
setup. \footnote{We use a split DZP basis set, a mesh cutoff of $200$ Ry, a Monkhorst-Pack $k$-point mesh of 1x2x1 and the LDA xc-functional\cite{PhysRevB.23.5048} to calculate the electronic structure. Supercell dimension($k_y$-points) of $ 27\text{\AA}\times 12.8\text{\AA}(101)$/$34.5\text{\AA}\times 25.5\text{\AA}(\Gamma)$/$49.3\text{\AA}\times 12.8\text{\AA}(\Gamma)$/$27\text{\AA}\times 17\text{\AA}(\Gamma)$ is used for inelastic transport for the pristine/SW/edge/adsorbate configuration.} We consider a suspended graphene sheet located $5$
\AA\ below the tip of a gold STM probe model. The ``left'' semi-infinite lead is attached
to the probe, while the ``right'' semi-infinite leads are attached at \emph{both} sides
of the graphene sheet. We consider a voltage bias between ``left'' and ``right''
leads. The electron-phonon coupling ($\mathbf M^\lambda$) is inherent to a coupling region (green $+$ black atoms in \Figref{Dynamic}) of phonon modes (index $\lambda$) calculated in a dynamical region (black atoms).  
Floating orbitals are included between the STM tip and the graphene
sample, to give a better description of the exponential decay of tunneling
conductance.\cite{Ghost}

Following the Lowest Order Expansion (LOE),\cite{LOE} simplified and efficient expressions
for the IETS signals can be derived under the assumption of weak electron-phonon
coupling. The LOE expressions involve just the evaluation of the spectral density matrices
for left/right moving states, $\mathbf A_{L/R}(\varepsilon)$, at the chemical potentials,
$\varepsilon=\mu_L,\mu_R$, corresponding to the threshold voltage bias ($V_b$) for
excitation of a given phonon ($\lambda$), $|\mu_L-\mu_R|=\hbar\omega_\lambda$.  Thus the
LOE expression does not \textit{per se} reflect changes in the DOS above the phonon excitation threshold. However, in the context of STS on gated graphene, this is highly relevant since the
behavior of the DOS leads to a distinct dip in the differential conductance at specific
applied voltage, $V_b=V_D$, \cite{Crommie} enabling a determination of the local chemical
potential of graphene.

In order to encompass this important variation in the DOS above threshold we make the
following observations (see also \Figref{phase}(a)). The expressions for the current which
gives rise to inelastic signals have a form exemplified by the inelastic contribution to
the current within LOE,
\begin{align}
  I_i &\approx \frac{e}{\hbar} \left( \coth \frac{\hbar \omega _\lambda}{2k_BT} -
    \coth  \frac{\hbar \omega _\lambda - eV_b}{2k_BT} \right)\nonumber 
  \\
  &\quad\times \int ^\infty _{-\infty}\!\!\!\! \mathrm d\varepsilon\,
  \Tr{ \mathbf M _\lambda \tilde{\mathbf A }_L (\varepsilon)\mathbf M _\lambda \mathbf A
      _R (\varepsilon -\hbar \omega _\lambda) }\nonumber
  \\
  &\quad\times \left\lbrace f_L(\varepsilon) -f_R (\varepsilon -\hbar \omega _\lambda)
  \right\rbrace ,\label{eq.Ii}
\end{align}
where $\tilde{\mathbf A }_L$ is the time reversed left spectral function, and $f_{L/R}$ is
the left/right occupation function.  Here the coth-terms yield the IETS-signal, namely a
sharp step in the differential conductance, $\partial_{V_b} I_i$, for
$V_b=\hbar\omega_\lambda$ at low temperature. Above threshold
($|V_b|>\hbar\omega_\lambda$) the step behavior is unimportant and we are left with the
bias-behavior of the integral. For finite bias both the filling of states ($f_{L/R}$) as
well as the states in the device, that is, the spectral functions, change with $V_b$.
However, in this STM setup the device is strongly coupled to the right lead (graphene) and
is very weakly coupled to the left lead (probe).
Consequently the potential in the device is pinned to that of the right lead, which is the Fermi
level $\mu_R=\varepsilon_F$ of the gated graphene lead. The DOS of the gold
STM probe varies slowly w.r.t. energy. Thus if we define $\mu_L=\varepsilon_F+eV_b$ with
the right chemical potential fixed at $\mu_R=\varepsilon_F$, the \emph{only} important
voltage dependent term is the Fermi-function, $f_L(\varepsilon)=n_F(\varepsilon-eV_b)$
inside the integral in \Eqref{eq.Ii} yielding an approximate $\delta$-function in the
differential conductance expression,
\begin{equation}
\partial _{V_b} I_i \approx \gamma _{i,\lambda} \partial _{V_b} \mathcal{I}^{\mathrm{sym}},
\label{dI1}
\end{equation}
where
\begin{equation}
\gamma _{i,\lambda}=\Tr{\mathbf M_\lambda \tilde{\mathbf A }_L(eV_b+\varepsilon _F) \mathbf M_\lambda \mathbf A_R (eV_b+\varepsilon _F -\hbar \omega _\lambda) },
\label{dI2}
\end{equation}
and $\mathcal{I}^{\mathrm{sym}}$ is a universal function. \cite{LOE} We are therefore
again left with evaluating the spectral functions at only two energies,
$\varepsilon=eV_b+\varepsilon_F$ and $\varepsilon=eV_b+\varepsilon_F-\hbar \omega
_\lambda$. Equation~\ref{dI1} is equivalent to the usual LOE expression but valid above
threshold due to the constant tip-DOS. The same argument can be applied to the other terms
in LOE.

In practice we calculate the LOE differential conductance for a range of $\mu_L$, see
\Figref{phase}(b), and can obtain the differential conductance by following the contour along
the $\mu_L = V_b$ direction (green line), and projecting it onto the $\mathrm dI/\mathrm
dV_b$, $V_b$ plane. Gating the graphene sheet corresponds to shifting the Fermi level by a
constant over all bias voltages, and can be obtained from \Figref{phase}(b) by a
translation of the contour along the chemical potential axis. \\

\begin{figure}[!h]
  \centering
  \includegraphics[width=0.5\textwidth]{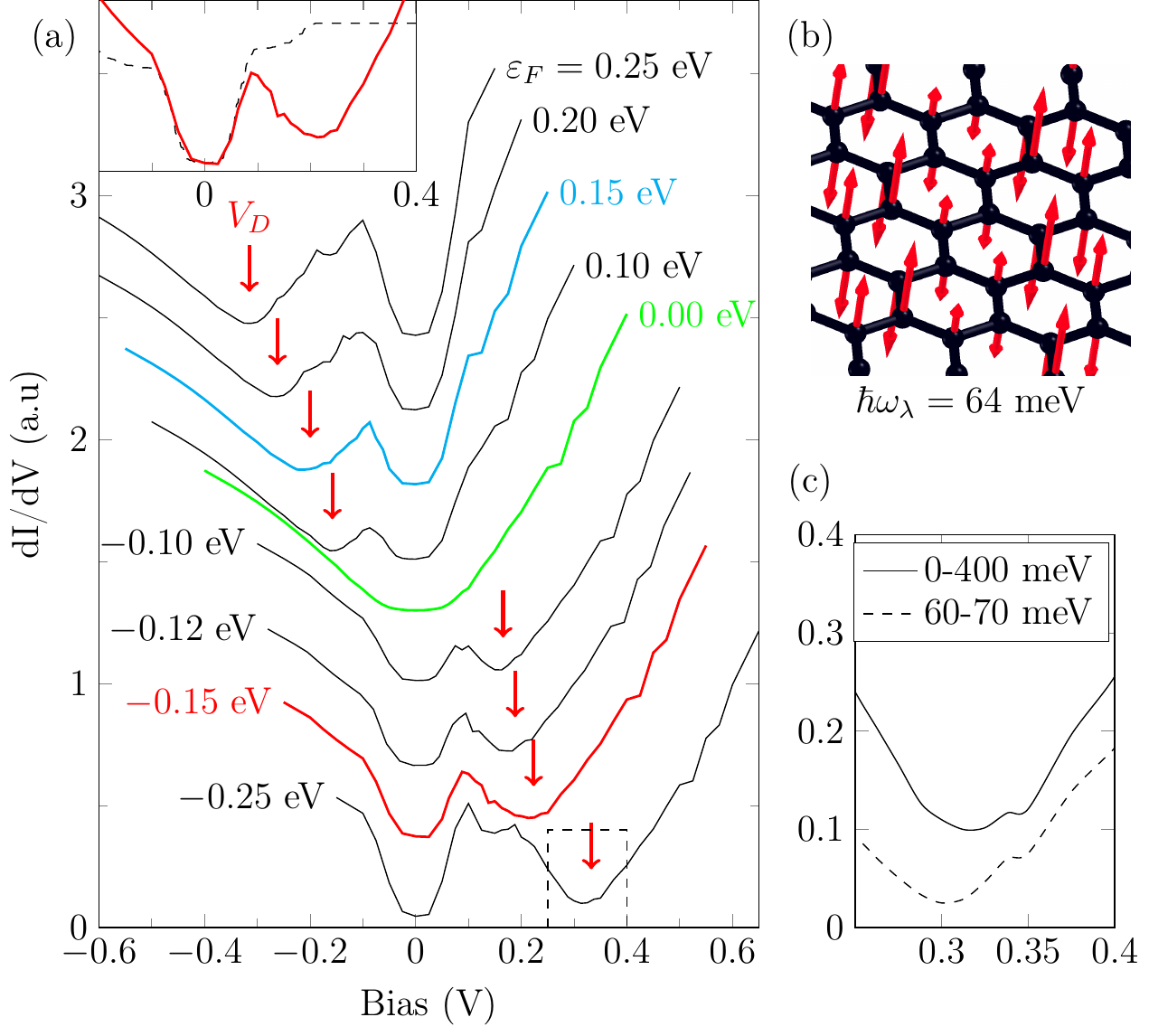} 
  \caption{(Color online)(a) Calculated STS spectra of pristine graphene at different Fermi levels. Colors represent the lines
      in \Figref{phase}(b). Inset: Comparison between STS spectra, at $\varepsilon_F=-0.15$ eV including/ignoring (red/black) DOS effects above threshold bias. (b) Out-of-plane acoustic graphene phonon. (c) Close-up of the dip at $V_D$ for $\varepsilon_F=-0.25$ eV, including phonons in
      different energy ranges.}
  \label{Gate}
\end{figure}
\noindent%
\textit{Results --- pristine graphene.}\quad %
Calculated STS spectra on pristine graphene for a number of different Fermi levels are shown in
\Figref{Gate}(a). The gap feature around $V_b=0$ of width $0.13$ V is reproduced in detail and the dip at $V_D$, caused by inelastic tunneling into the charge neutrality point of graphene, appears outside the gap as seen in experiments.\cite{Crommie,BN,PhysRevLett.104.036805} As the gate is applied, $V_D$ moves across the spectrum changing polarity while the position and
width of the gap feature is stable. The inset shows how significant the difference in results are when ignoring/including above threshold terms. It also shows how the usual LOE calculation works well
below threshold and captures the inelastic steps.\\
Most major steps in differential conductance come from acoustic out-of-plane phonons at energies just below $67$ meV.\cite{2007} In particular the mode shown in \Figref{Gate}(b) gives a large contribution.  
However we find that acoustic out-of-plane graphene phonons with energies as
low as $42$ meV give considerable contributions as well. We also find important inelastic signals from optical graphene phonons at energies above $67$ meV. The additional features away from $67$ mV make up about half the signal, and have not been included in previous
studies.\cite{Wehling} If we restrict our calculations to phonons in the $60$--$70$ meV
range, we obtain a $15$ mV change in ($V_D$), see \Figref{Gate}(c), and changes in both the width and height of the inelastic gap. The change in $V_D$ is caused mostly by the experimentally observed\cite{PhysRevLett.104.036805} inelastic signal near $150$ mV, coming from the optical in-plane modes, and occurs for $|V_D|>150$ mV. 
In STS experiments $V_D$ is used to extract the energy position of the charge neutrality point from $E_D=e|V_D|-\hbar \omega_0$ where $\hbar \omega_0=63$ meV is half the width of the gap feature which corresponds to the energy of an acoustic out-of-plane graphene phonon.\cite{Crommie,BN} 
The change in $V_D$ could explain why all points with $|E_D|<100$ meV in the $E_D$ vs gate voltage plot of Ref.\citenum{BN} fall below the fitted line. 
Local charge-carrier density ($n$) of graphene is also extracted from $V_D$ in STS experiments.\cite{BN,nl2041673,nl3039508}  
Mistaking $E_D=100$ meV for $E_D=115$ meV results in a 32\% error in $n$.
To capture these experimental details one must include
several phonons, and account for their impact in an \emph{ab initio} manner.\\
Encouraged by the agreement for pristine graphene we
next predict the inelastic signals from various defects to shed light on what information
can be obtained from STM-IETS.\\
\begin{figure}[!h]
  \centering
  \includegraphics[width=0.45\textwidth]{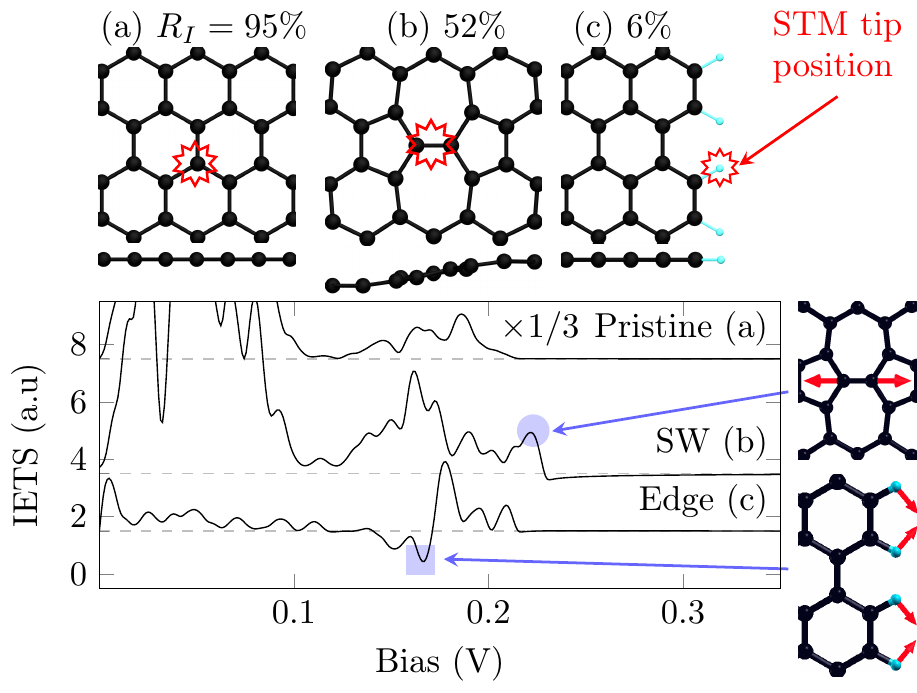} 
  \caption{IETS as a function of bias for pristine graphene, a Stone-Wales defect, and a
      hydrogen passivated armchair edge (geometries shown above plot). The blue marker
      indicate characteristic signals. The fraction of differential conductance coming from the inelastic channel ($R_I$) is shown above the geometries.}
  \label{defect}
\end{figure}

\noindent%
\textit{Results --- Structurally defected graphene.}\quad %
In \Figref{defect} we show the calculated IETS spectra from an on-top position in pristine
graphene (a), directly above a Stone-Wales defect (SW) (b), and above a passivated
armchair edge (c). The result shown for pristine graphene is the same at hollow sites and
bridge sites. 
We find that the gap feature and low voltage IETS above a SW is very similar
to that of pristine graphene. The gap has also been observed experimentally for regions
with heptagon-pentagon defects.\cite{GrainBoundary} 
However, a characteristic signal can be seen at  $V_b=223$ mV bias, above any of the pristine graphene phonon bands which can be traced to the high frequency stretch mode localized at the
twisted C-C bond shown in \Figref{defect}.\\
Ignoring the out-of-plane buckling introduced to the graphene sheet near a SW, and calculating the IETS for a flat SW system, leads to a $5$ mV blue-shift of the signal from the twisted C-C bond as previously proposed.\cite{Sine} We also see strong signals at low bias. These signals are caused by low-frequency sine-like out-of-plane modes. These modes couple strongly to the current because they break the mirror symmetry across the twisted C-C bond.
In the buckled system, this symmetry is inherently broken, leading to an increase in elastic tunneling.
Measuring strong low bias inelastic signals and a $228$ mV signal above a SW therefore indicates that it is in a metastable flat configuration, whereas increased elastic transmission and a $223$ mV signal is a sign of local buckling. Above a passivated armchair edge a dip in IETS is seen at $V_b=168$ mV caused by a collective transverse mode of the Hydrogen atoms shown in \Figref{defect}. 
Changing the mass of the passivating agent to that of Fluor, we observe a corresponding change in the position of the inelastic signals. This indicates that IETS can be used to obtain knowledge of graphene edge passivation.\\

\begin{figure}[!h]
  \centering
  \includegraphics[width=0.4\textwidth]{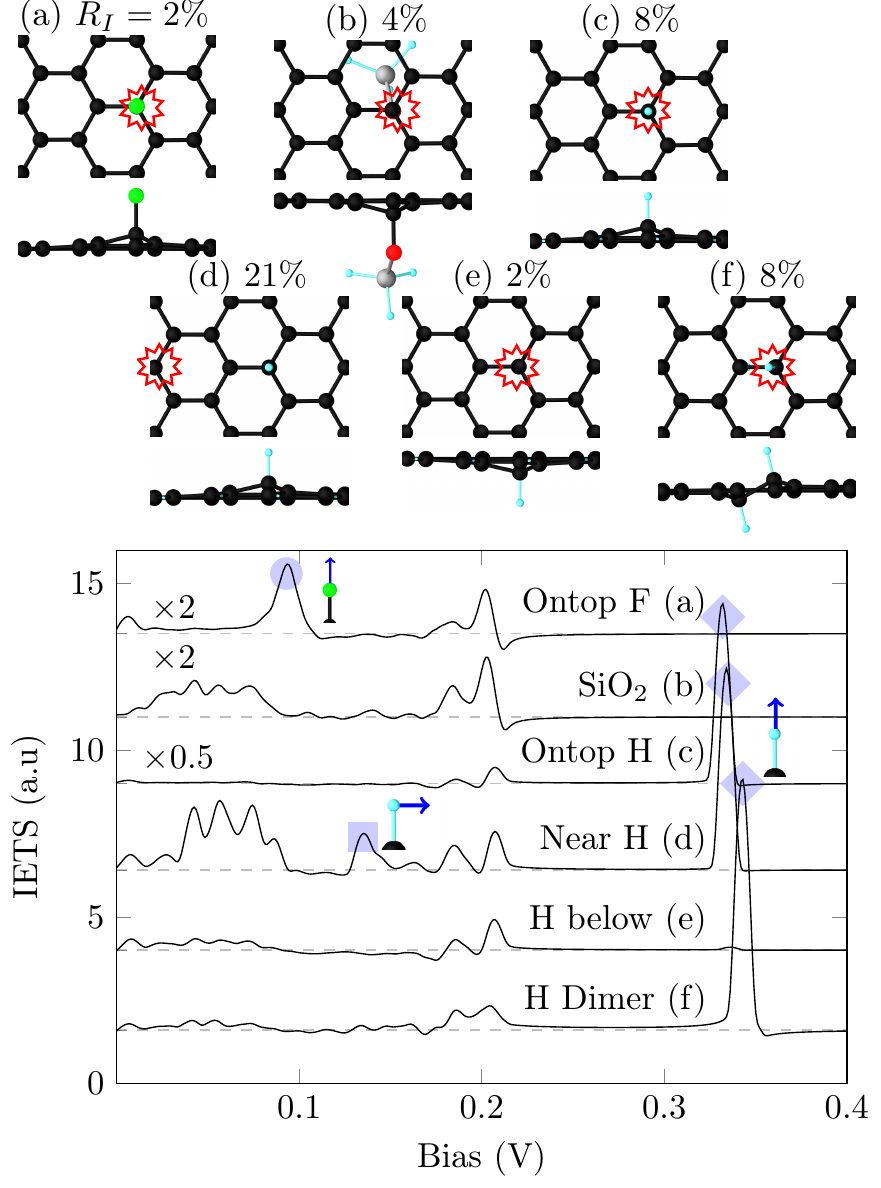} 
  \caption{IETS as a function of bias voltage, for various adsorbates on graphene
      (geometries shown above plot). Fingerprints for each adsorbate
      is marked and the phonon shown.The fraction of differential conductance coming from the inelastic channel ($R_I$) is shown above the geometries.}
  \label{Adsorbates}
\end{figure}
\noindent%
\textit{Results --- Adsorbates on graphene.}\quad %
In \Figref{Adsorbates}  we show IETS spectra from a range of different covalently bonded
impurities (Fluor, Hydrogen) and a model of strong interaction to a SiO$_2$ substrate.  In \Figref{Adsorbates}(a) a
clear inelastic signal from the longitudinal mode of a Flour adsorbate is seen at $95$ mV. Above a Hydrogen adsorbate, we
see a strong inelastic peak at $332$ mV caused by the stretch mode of the C-H
bond (see \Figref{Adsorbates}(c)). This signal serves as a fingerprint for a Hydrogen impurity above the graphene sheet as opposed to below where the signal disappears as can be seen in \Figref{Adsorbates}(e). The corresponding STS spectra show a strong zero-energy peak\cite{Hydrogen}, this behavior is however expected for all covalently bonded impurities,\cite{Midgap} above or below the sheet, and can therefore not be used as a fingerprint.\\
The STS spectra on the hydrogenated system with the probe above a Carbon atom $4.25$
\AA\ laterally away from the impurity in \Figref{Adsorbates}(d), shows additional signals.
The graphene out-of-plane phonon signals reappear and a signal is also seen at $134$ mV
caused by a transverse mode of the C-H bond.  
Above a graphane-like hydrogen dimer \Figref{Adsorbates}(f) the signal caused by the C-H bond stretch mode is seen,
however here it is caused by two degenerate modes and blue-shifted by 11--16 mV indicating a lower energy configuration.\\
Common for all the imperfect systems is that the gap seen in pristine graphene is
quenched as indicated by the severe reduction of the inelastic conductance ratio ($R_I$) in \Figref{defect} and \ref{Adsorbates}. Out-of-plane corrugations in the graphene sheet can lift the suppression of
elastic tunneling if they are on the same length scale as the graphene lattice constant.\cite{Wehling} 
Our results indicate that defects can also lift the suppression locally. This is because the selection rules causing the suppression
in pristine graphene is a result of the translational symmetry of the crystal lattice. When this symmetry
is broken the suppression is lifted and the elastic tunneling dominates. The expected
order of magnitude change in tunneling conductance should lead to bright spots in STM
topographies. In the case of granular CVD graphene protruding grain boundaries are often
attributed to localized electronic states.\cite{GrainBoundary} Our results here point out
that one may expect increased tunneling near disordered areas of graphene, even if no
localized electronic states are present and the area is completely flat.
As seen in \Figref{Adsorbates}(b)(e) this is also the case for strong interaction with a SiO$_2$ substrate or adsorbates sitting below the graphene sheet, which should therefore be visible as protruding from the graphene sheet.\\

In summary, we have presented a first principles method and used it for calculations of
IETS and STS spectra of pristine and defected graphene. We showed how measured STS spectra
on pristine gated graphene can be reproduced in detail as a function of gating. The
inclusion of several phonons had a strong impact on all aspects of the STS spectrum of
pristine graphene. In particular we found that including optical in-plane phonons changed
the $V_D$ value for certain gate voltages. This is of importance for studies where IETS
is used to probe the local doping of graphene\cite{BN,nl2041673,nl3039508} where it may lead to a
significant overestimation of the local charge inhomogeneity.  We predicted the IETS of
typical imperfections in graphene, and demonstrated how these can
yield characteristic fingerprints revealing e.g. adsorbate species or local buckling.
Additional elastic contributions above defects should make them protrude in STM regardless of actual geometric or electronic
structure and care is needed when interpreting STM images.

We gratefully acknowledge discussions with Thomas Frederiksen, Aran Garcia-Lekue and Rasmus Bjerregaard Christensen.
%

\end{document}